\documentclass[preprint,showpacs,amsmath,amssymb,floatfix,pre]{revtex4}
\usepackage{graphicx}
\begin{document}
\title{The Ising antiferromagnet on an anisotropic simple cubic lattice in the presence of a magnetic field}
\author{Octavio D. R. Salmon}
\email{octaviors@gmail.com}
\author{Minos A. Neto}
\email{minosneto@pq.cnpq.br}
\author{J. Roberto Viana}
\email{vianafisica@bol.com.br}
\author{Igor T. Padilha}
\email{igorfis@ufam.edu.br}
\affiliation{Departamento de F\'{\i}sica, Universidade Federal do Amazonas, 3000, Japiim,
69077-000, Manaus-AM, Brazil}

\author{J. Ricardo de Sousa}
\email{jsousa@edu.ufam.br}
\affiliation{Departamento de F\'{\i}sica, Universidade Federal do Amazonas, 3000, Japiim,
69077-000, Manaus-AM, Brazil}
\affiliation{National Institute of Science and Technology for Complex Systems, 3000, Japiim,
69077-000, Manaus-AM, Brazil}
\date{\today}

\begin{abstract}

We have studied the anisotropic three-dimensional nearest-neighbor Ising model with competitive interactions in an 
uniform longitudinal magnetic field $H$. The model consists of ferromagnetic interaction $J_{x}(J_{z})$ in the $x(z)$ direction and 
antiferromagnetic interaction $J_{y}$ in the $y$ direction. We have compared our calculations within a effective-field theory in clusters with 
four spins (EFT-4) in the simple cubic (sc) lattice with traditional Monte Carlo (MC) simulations. The phase diagrams in the 
$h-k_{B}T/J_{x}$ plane ($h=H/J_{x}$) were obtained for the particular case $\lambda_{1}=J_{y}/J_{x} (\lambda_{2}=J_{z}/J_{x})=1$ (anisotropic sc). Our results 
indicate  second-order frontiers for all values of $H$ for the particular case $\lambda_{2}=0$ (square lattice), while in  
case  $\lambda_{1}=\lambda_{2}=1$, we observe first- and second-order phase transitions in the low and high temperature limits, 
respectively, with presence of a tricritical point. Using EFT-4, a reentrant behavior at low temperature was observed in contrast with 
results of MC.   

\end{abstract}

\maketitle

\section{Introduction\protect\nolinebreak}

In recent years, the effect of a longitudinal field in the Ising antiferromagnet on an anisotropic simple cubic (sc) lattice has been 
discussed. The experimental example is compound $(C_{2}H_{5}NH_{3})_{2} CuCl_{4}$ \cite{dejongh}. The differential magnetic susceptibility 
$\chi=\left(\frac{\partial M}{\partial H}\right)_{T}$ of this compound  was analysed as a function of an extra external field ($0-2$ kOe) 
and of temperature ($1-30$ K; $T_{c}=10.20$ K). The compound is a typical layer-type ferromagnet, with a very weak antiferromagnetic coupling 
between the $Cu^{2+}$ layers, where has been established the magnetic phase diagram of the antiferromagnetic structure. One of the attractive 
points of investigating the properties of $(C_{2}H_{5}NH_{3})_{2} CuCl_{4}$ is that as a consequence of the antiferromagnetic interlayer coupling 
we may obtain quantitative information about the anisotropy and $J_{AF}$ (exchange coupling) by investigating the field dependence of the susceptibility 
at $T<T_{c}$. In previous papers \cite{dejongh2,bloembergen,dejongh3,miedema} it has been reported that the $Cu$ compounds of general formula 
$(C_{n}H_{2n+1}NH_{3}CuX_{4})$, where $n=1$, $2$, $3$, $4$, $5$, $6$, $10$ and $X=Cl$ or $Br$, may be considered as consisting of nearly isolated 
magnetic layers. Other example of the compound with cubic anisotropy are antiferromagnet $K_{2}MnF_{4}$ \cite{mulder}, $(CH_{3}NH_{3})_{2}MnCl_{4}$ 
and $(CD_{3}ND_{3})_{2}MnCl_{4}$ \cite{heger}.

Three dimensional (3D) Ising models and (pseudo-) Ising physical systems have been analysed extensively \cite{domb,wolf}. Graim and Landau \cite{graim} 
 studied the critical behavior of a spin-$1/2$ Ising model on a simple cubic lattice with spatially anisotropic nearest-neighbor 
coupling using the Monte Carlo method. This model on an anisotropic square lattice was investigated by using a modified mean-field theory in which the 
intrachain is treated exactly and the interactions between chains are considered in the mean-field theory \cite{stout, hone, sato}. Various 
approximate methods have show this critical bahavior of the curve $T_{N}$ versus $H$, such as mean field approximation (MFA) \cite{garrett,ziman}, 
effective-field theory (EFT) \cite{zukovic}, mean field renormalization group (MFRG) \cite{slotte}, effective-field renormalization group (EFRG) 
\cite{neto2004}, Monte Carlo simulation (MC) \cite{landau,landau1976,ferrenberg}, and high-temperature series expansion (SE) \cite{bienenstock}. For the 
case of the 3D lattice, the theoretical calculations show disagreement between differente methods. The results obtained by the MFA and EFT methods show 
a reentrant behavior in the phase diagram in low-temperature, i.e., if $H$ is just above $H_{c}$, then these are two phase transitions as the 
temperature is increased.         

In recent years, the effect a longitudinal field in the Ising antiferromagnetic on an anisotropic square lattice was explored by MC \cite{viana2009}. 
Although MC simulations play an important role for the study of phase transitions and critical phenomena, the well-known difficulties arise when one uses standard algorithms (one-flip algorithms) \cite{metropolis} for the study of first-and second-order phase transitions. This has contributed for the development of alternative MC methods, such as parallel-tempering \cite{nemoto}, cluster algorithms \cite{wolff}, multicanonical algorithms \cite{berg}, and more recently Wang-Landau method \cite{wanglandau}.

In the present paper we use the MC simulations and effective-field theory in clusters with four spins (EFT-4). We investigate the first- and second-order phase transition in the plane $h-k_{B}T/J_{x}$ of the Ising superantiferromagnet on an anisotropic simple cubic lattice in the presence of a magnetic field. Standard finite-size scaling techniques were used to  estimate the critical temperatures. In Section II we present the model and formalism. The numerical 
results and discussions are given in Section III. Finally, the last section is devoted to conclusions.

\section{Model and Formalism}
\subsection{Hamiltonian}

\begin{figure}[htbp]
\centering
\includegraphics[width=7.0cm,height=7.0cm]{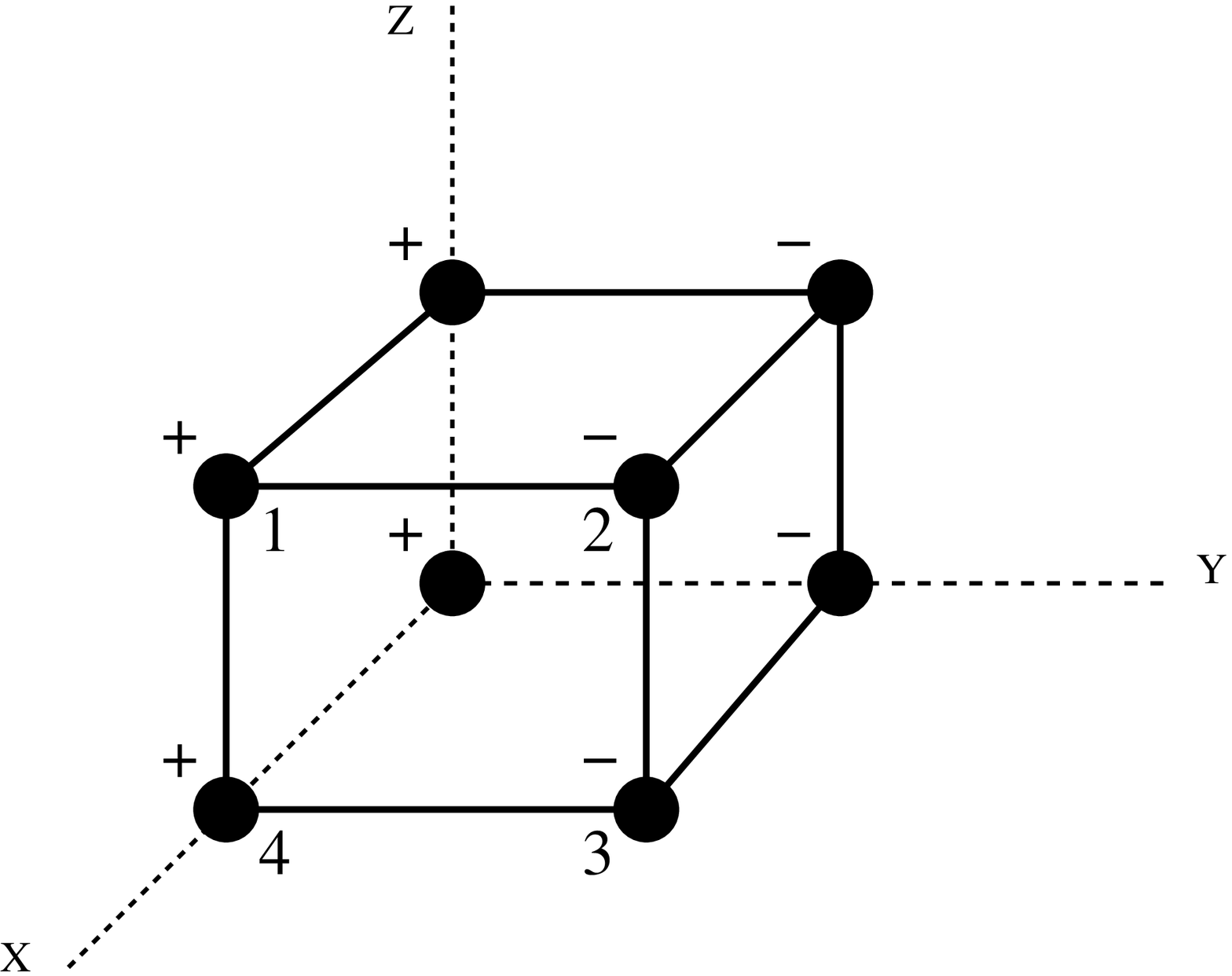}
\caption{Configuration of the superantiferromagnetic system of spins used to treat the model whose hamiltonian is described in Eq. (\ref{1}).} 
\label{cubo2}
\end{figure}

The model in this work is the nearest-neighbor ($nn$) Ising antiferromagnetic in a longitudinal field magnetic divided into two equivalent interpenetrating 
sublattices $A$ e $B$, that is described by following Hamiltonian

\begin{eqnarray}
\mathcal{H} =
-J_{x}\sum\limits_{i,\overrightarrow{\delta_{x}}}\sigma _{i}^{z}\sigma_{i+\overrightarrow{\delta_{x}}}^{z}
+J_{y}\sum\limits_{i,\overrightarrow{\delta_{y}}}\sigma _{i}^{z}\sigma_{i+\overrightarrow{\delta_{y}}}^{z} 
-J_{z}\sum\limits_{i,\overrightarrow{\delta_{z}}}\sigma _{i}^{z}\sigma_{i+\overrightarrow{\delta_{z}}}^{z}
-H\sum\limits_{i}\sigma_{i}^{z}, 
\label{1}
\end{eqnarray}%
where $\sigma_{i}^{\mu}$ is the $\mu(=x,y,z)$ component spin-$1/2$ Pauli operator at site $i$, $J_{x}(J_{y},J_{z})$ is the exchange coupling along the $x(y,z)$ 
axis, $\delta_{x}(\delta_{y},\delta_{z})$ denotes the nearest-neighbor vector along the $x(y,z)$ axis and $H$ is the longitudinal magnetic field. We define the parameters $\lambda_{1}=J_{y}/J_{x}$ and $\lambda_{2}=J_{z}/J_{x}$. The ordered state for low temperatures and fields is the superantiferromagnetic order 
(SAF), which  is characterrized by a parallel spin orientation in horizontal direction and an antiparallel spin orientation of a parallel spin orientation of 
nearest-neighbors in vertical direction and therefore exhibit N\'eel order within the initial sublattice $A$ and $B$ (see figure (\ref{cubo2})). 

If $\lambda_{2}=0$ ($J_{z}=0$), the lattice is now composed of independent planes, so the model is exactly solved for $H=0$, and the critical temperature is 
obtained by the following relation \cite{onsager} 

\begin{equation}
\sinh\left(\frac{2J_{x}}{k_{B}T_{N}}\right)\sinh\left(\frac{2J_{y}}{k_{B}T_{N}}\right)=1,
\label{2} 
\end{equation}

where for the particular isotropic case $J_{x}=J_{y}=J$ ($\lambda_{1}=1$ )  we have  $k_{B}T_{N}/J=2/\ln(1+\sqrt2)$. For $H \neq 0$, with $\lambda_{1} \neq 1 $, $\lambda_{1}=1$ we have an Ising model with an external magnetic field on an anisotropic square lattic (\ref{1}), which  was already studied by MC \cite{viana2009}. Accordingly, we improve the understanding of this model by studying it by means of the Effective-Field Theory and Monte Carlo simulations for the case 
$\lambda_{1}=\lambda_{2}$ and $H \neq 0$. 

\subsection{Monte Carlo Simulation}

In order to implement the present model to perform MC by the Metropolis Algorithm, the simple cubic lattice of size $L$ having $L\times L\times L$ sites is decomposed into two sublattices ($A$ and $B$) with opposite spins, corresponding to the SAF ground state. To meassure the SAF order, the appropriate order parameter is defined by $\left\langle m_{s}\right\rangle=\left\langle(m_{A}-m_{B})/2\right\rangle$, where $\left\langle m_{\mu}\right\rangle=\left\langle\frac{2}{N}\sum_{i\in \mu}\sigma_{i}\right\rangle$ is the magnetization of the sublattice $\mu=A$, $B$ and $N=L^{3}$ number of spins. The susceptibility related to this order parameter is defined as follows:

\begin{equation}
\chi = L^{d}(\langle m_{s}^{2}\rangle - {\langle m_{s}\rangle}^{2})/T, 
\end{equation}
where $d$ is the dimension of the lattice. 

In our simulations we have considered lattices with periodic boundary conditions. In order to determine the system's behavior in the thermodynamic limit ($L\rightarrow\infty$), which is imposible to implement on account of computational limitation, we have to  use a finite-size scaling theory \cite{fisher}. Accordingly, to locate the critical temperature for second-order phase transitions, we  approximately locate the crossing point of curves for different sizes of the fourth-order cumulant $U_{4}(L)$ (Binder Cumulant) defined as \cite{binder}

\begin{equation}
U_{4}(L)=1-\frac{\left\langle m_{s}^{4}\right\rangle}{3\left\langle m_{s}^{2}\right\rangle^{2}},
\label{3}
\end{equation}
where $\left\langle m_{s}^{2}\right\rangle$ and $\left\langle m_{s}^{4}\right\rangle$ are the canonical averages of the second and fourth moments of magnetization, respectively. On the other hand, the critical temperature $T_{c}$ can also be  obtained by means of the relation  $T_{c}^{L}=T_{c}+aL^{-1/\nu}$, where $T_{c}^{L}$ is the pseudocritical temperature corresponding to the size $L$, and $\nu$ is the correlation critical exponent. For first-order phase transitions the same formula applies by setting $1/\nu = d$. A more extensive description of the Monte Carlo method was published elsewhere \cite{landau1976} and the reader is referred there for further details. We performed simulations for $\lambda_{1}=\lambda_{2}$, for several values of $h$, with a temperature step $\Delta T=0.01$ and runs comprising up to $2\times10^{5}$ MC steps after equilibration. The statistical errors of the MC simulations used for the estimation of $T_{N}(\lambda_{1},\lambda_{2},h)$ of a particular $\lambda_{1}$, $\lambda_{2}$ and $h=H/J_{x}$ were found much smaller than the statistical errors coming from the fact that we used. Therefore, the error bars are not shown in our graphs because they are smaller than the symbol sizes.

\subsection{Effective-Field Theory}

As a starting point, the averages of a general function involving spin operator components $O(\{n\})$ are obtained by \cite{denise}

\begin{equation}
\left\langle \mathcal{O}(\{n\})\right\rangle =\left\langle \frac{\text{Tr}%
{\{n\}}\mathcal{O}(\{n\})e^{-\beta \mathcal{H}_{\{n\}}}}{%
Tr_{\{n\}}e^{-\beta \mathcal{H}_{\{n\}}}}\right\rangle ,  
\label{4}
\end{equation}%
where the partial trace Tr$_{\{n\}}$ is taken over the set $\{n\}$ of spin variables (finite cluster) specified by the multisite spin Hamiltonian $H_{\{n\}}$ and $\left\langle \cdot 
\cdot \cdot \right\rangle $ indicates the usual canonical thermal average.

The method treats the effects of the surrounding spins of a finite cluster with $N$ spins through a convenient differential operator technique \cite{honmura} such that, in contrast to 
the usual MFA procedure, all relevant self-spin correlations are taken exactly into account. The interactions within the cluster are exactly treated and the effect of the remaining 
lattice spins is treated by a given approximation (here we use the random phase approximation-RPA).

To treat the model (\ref{1}) by the EFT approach, we consider a simple example in cluster of size $N=4$ spins, and the Hamiltonian for this 
cluster is given by

\begin{equation}
-\beta\mathcal{H}_{4}=-K\lambda_{1}\sigma^{z}_{1}\sigma^{z}_{2}+K\lambda_{2}\sigma^{z}_{2}\sigma^{z}_{3}-K\lambda_{1}\sigma^{z}_{3}\sigma^{z}_{4}
+K\lambda_{2}\sigma^{z}_{4}\sigma^{z}_{1}+\sum_{r=1}^{4}a_{r}\sigma^{z}_{r},
\label{5}
\end{equation}%
where $a_{r}=L-K\sum_{\delta_{r}}\sigma_{r+\delta_{r}^{z}}$ with $K=\beta J_{x}$, $\lambda_{1,2}=J_{y,z}/J_{x}$, $L=\beta H$ and $\delta_{r}$ corresponds 
to $nn$ vectors.

Substituting Eq. (\ref{5}) in (\ref{4}), we obtain the average magnetizations in sublattices $A$ and $B$, respectively, by

\begin{equation}
m_{A}=\left\langle \sigma _{1}^{z}\right\rangle =\left\langle \frac{\partial \ln \mathcal{Z}_{4}(\mathbf{a})}{\partial (\beta a_{1})}\right\rangle ,
\label{6}
\end{equation}%
and%
\begin{equation}
m_{B}=\left\langle \sigma _{2}^{z}\right\rangle =\left\langle \frac{\partial \ln \mathcal{Z}_{4}(\mathbf{a})}{\partial (\beta a_{2})}\right\rangle ,
\label{7}
\end{equation}%
with%
\begin{equation}
\mathcal{Z}_{4}(\mathbf{a})=\text{Tr}_{\{\mathbf{\sigma }\}}e^{-\beta \mathcal{H}_{4}},  
\label{8}
\end{equation}%
where $\textbf{a}=(a_{1},a_{2},a_{3},a_{4})$ and $\{\sigma \}=\{\sigma_{1}^{z},\sigma_{2}^{z},\sigma_{3}^{z},\sigma_{4}^{z}\}$.

Using the identity $\exp(\textbf{a}\cdot \textbf{D})f(x)=f(x+a)$, where $\textbf{D}=(D_{1},D_{2},D_{3},D_{4})$ and $\textbf{x}=(x_{1},x_{2},x_{3},x_{4})$ are four-
dimensional differential operator and vector, respectively, $D_{\mu }=\frac{\partial }{\partial x_{\mu }}$, and also the van der Waerden identity for $\sigma _{i}^{z}$ component 
Pauli spin operator, i.e., $\exp (\lambda\sigma _{i}^{z})=\cosh (\lambda )+\sigma _{i}^{z}\sinh (\lambda )$, Eqs. (\ref{6}) and (\ref{7}) are rewritten 
as ($\mu =A$ or $B$)

\begin{equation}
\begin{array}{c}
m_{\mu }=\left\langle \prod\limits_{\mathbf{\delta }_{1}}^{z-2}\left(
\alpha _{1}+\sigma_{1A+\mathbf{\delta}_{1}}^{z}\beta _{1}\right)
\prod\limits_{\mathbf{\delta}_{2}}^{z-2}\left(\alpha _{2}+\sigma _{2B+%
\mathbf{\delta }_{2}}^{z}\beta _{2}\right) \right. \\ 
\left. \prod\limits_{\mathbf{\delta }_{3}}^{z-2}\left(\alpha _{3}+\sigma
_{3B+\mathbf{\delta}_{3}}^{z}\beta _{3}\right) \prod\limits_{\mathbf{%
\delta }_{4}}^{z-2}\left( \alpha _{4}+\sigma_{4A+\mathbf{\delta}_{4}}^{z}\beta
_{4}\right) \right\rangle \left. f_{\mu }(\mathbf{x+L})\right| _{\mathbf{x}%
=0}%
\end{array}
\label{9}
\end{equation}%
with%
\begin{equation}
f_{A}(\mathbf{x})=\frac{\partial \ln \mathcal{Z}_{4}(\mathbf{x})}{\partial
x_{1}}=\frac{\psi_{1}+\psi_{2}+\psi_{3}+\psi_{4}}{\phi_{1}+\phi_{2}+\phi_{3}+\phi_{4}},  
\label{10}
\end{equation}%
and 
\begin{equation}
f_{B}(\mathbf{x})=\frac{\partial \ln \mathcal{Z}_{4}(\mathbf{x})}{\partial
x_{2}}=\frac{\psi_{1}-\psi_{2}+\psi_{3}-\psi_{4}}{\phi_{1}+\phi_{2}+\phi_{3}+\phi_{4}},  
\label{11}
\end{equation}%
where $\psi_{1}=\sinh\left(C_{1}\right)+e^{2K\left(\lambda_{1}+\lambda_{2}\right)}\sinh\left(C_{2}\right)$, 
$\psi_{2}=\sinh\left(C_{3}\right)+e^{2K\left(\lambda_{1}-\lambda_{2}\right)}\sinh\left(C_{4}\right)$,  
$\psi_{3}=\sinh\left(C_{5}\right)+e^{2K\left(-\lambda_{1}+\lambda_{2}\right)}\sinh\left(C_{6}\right)$, 
$\psi_{4}=\sinh\left(C_{7}\right)+e^{-2K\left(\lambda_{1}+\lambda_{2}\right)}\sinh\left(C_{8}\right)$, 
$\phi_{1}=\cosh\left(C_{1}\right)+e^{2K\left(\lambda_{1}+\lambda_{2}\right)}\cosh\left(C_{2}\right)$, 
$\phi_{2}=\cosh\left(C_{3}\right)+e^{2K\left(\lambda_{1}-\lambda_{2}\right)}\cosh\left(C_{4}\right)$, 
$\phi_{3}=\cosh\left(C_{5}\right)+e^{2K\left(-\lambda_{1}+\lambda_{2}\right)}\cosh\left(C_{6}\right)$,   
$\phi_{4}=\cosh\left(C_{7}\right)+e^{-2K\left(\lambda_{1}+\lambda_{2}\right)}\cosh\left(C_{8}\right)$, 
$C_{1}=x_{1}+x_{2}+x_{3}+x_{4}$, $C_{2}=x_{1}-x_{2}-x_{3}+x_{4}$, $C_{3}=x_{1}+x_{2}-x_{3}+x_{4}$, 
$C_{4}=x_{1}-x_{2}+x_{3}+x_{4}$, $C_{5}=x_{1}-x_{2}+x_{3}-x_{4}$, $C_{6}=x_{1}+x_{2}+x_{3}-x_{4}$, 
$C_{7}=x_{1}-x_{2}-x_{3}-x_{4}$ and $C_{8}=x_{1}+x_{2}-x_{3}-x_{4}$.

The magnetization $m_{A}$ in Eq. (\ref{9}) is expressed in terms of multiple spin correlation functions. The problem becomes unmanageable when we 
try to treat exactly all boundary spin-spin correlation function present in Eq. (\ref{9}). Here we use a decoupling procedure that ignores all higher-order spin correlations on both right-hand sides in Eq. (\ref{9}), namely

\begin{equation}
\left\langle\sigma_{iA}^{z}\sigma_{jB}^{z}\dots\sigma_{lA}^{z}\right\rangle\simeq m_{A}m_{B}\dots m_{A},  
\label{12}
\end{equation}%
where $i\neq j\neq \dots \neq l$ and $m_{\mu }=\left\langle \sigma _{i\mu}^{z}\right\rangle $ $(\mu=A,B)$. The approximation (\ref{12}) neglects correlations 
between different spins but takes relations such as $\left\langle \left(\sigma_{i\nu }^{z}\right)^{2}\right\rangle =1$ exactly into account, while in the usual MFA 
all the self- and multi spin correlations are neglected. We can then rewrite the Eq. (\ref{9}) in the form

\begin{equation}
m_{A}=\left(\alpha_{1}+m_{A}\beta_{1}\right)^{2}\left(\alpha_{2}-m_{B}\beta_{2}\right)^{2}
\left(\alpha_{3}-m_{B}\beta_{3}\right)^{2}\left(\alpha_{4}+m_{A}\beta_{4}\right)^{2}
\left.f_{A}(\mathbf{x})\right|_{\mathbf{x}=0},  
\label{13}
\end{equation}%
and the expression for the magnetization in sublattice $B$ is given by

\begin{equation}
m_{B}=\left(\alpha_{1}+m_{B}\beta_{1}\right)^{2}\left(\alpha_{2}-m_{A}\beta_{2}\right)^{2}
\left(\alpha_{3}-m_{A}\beta_{3}\right)^{2}\left(\alpha_{4}+m_{B}\beta_{4}\right)^{2}
\left.f_{B}(\mathbf{x})\right|_{\mathbf{x}=0}.  
\label{14}
\end{equation}

Defining the uniform $m=\frac{1}{2}(m_{A}+m_{B})$ and staggered $m_{s}=\frac{1}{2}(m_{A}-m_{B})$ magnetizations, and using the identity 
$\exp\left(\textbf{a}\cdot\textbf{D}\right)F(\textbf{x})|_{\textbf{x}}=F(\textbf{a})$, we obtain 

\begin{equation}
m_{s}=\Lambda(m_{s},m,T,H)=\sum_{r=0}^{3}A_{2r+1}(m,T,H)m_{s}^{2r+1}
\label{15}
\end{equation}%
and

\begin{equation}
m=\sum_{r=0}^{4}B_{2r}(m,T,H)m_{s}^{2r},
\label{16}
\end{equation}%
where the expressions for the coefficients $A_{p}(m,T,H)$ and $B_{p}(m,T,H)$ are again omitted here.

We note that is not possible to calculate the firts-order transition line on the basis of only the equation of state, Eq. (\ref{15}), to solve this problem one needs 
to calculate the free energy for each state (P and SAF). Assuming that this equantion of state is obtained by the minimization of a given free energy functional like 
$\Phi(m_{s})$ (i.e., $\delta\Phi=0$), then after intergration we obtain 

\begin{equation}
\Phi(m_{s})=\Delta_{1}(T,H)+\Delta_{2}(T,H)\left[\frac{m_{2}^{2}}{2}-\sum_{r=0}^{3}A_{2r+1}(m,T,H)\frac{m_{s}^{2r+1}}{2r+2}\right], 
\label{17}
\end{equation}%
where $\Delta_{1,2}(T,H)$ are arbitrary functions which turn out to be irrelevant for searching the second and first-order transitions.

To obtain this phase transitions we use Maxwell construction, that correspond to the intersection point where the free energy between the phases are equal. In the case of the transitions between the SAF ordered and P disordered $(m_{s})$ phases we obtain the point of intersection $\Phi_{SAF}(m_{s})=\Phi_{P}(m_{s})$ from Eq. (\ref{17}), i.e, 
\begin{equation}
\sum_{r=0}^{3}A_{2r+1}(m,T,H)\frac{m_{s}^{2r}}{r+1}=1.
\label{18}
\end{equation}

The phase transition temperatures between the P and SAF states are found by simultaneously solving three transcendental expressions, Eqs. (\ref{15}), (\ref{16}) and (\ref{18}). For a second-order transition, we obtain $m_{s}=0$, while the first-order transition we have $m_{s}\neq0$, where this value corresponds to the discontinuity of the staggered magnetization at $T=T_{c}^{*}(H)$.

\section{Results and Discussion}

We obtained the phase diagrams in the $h-k_{B}T/J_{x}$ plane for $\lambda_{1}=\lambda_{2}=1$ of the model \ref{1} by using EFT-4 and MC simulatons.  Both methods confirmed the existence of a tricritical point in a critical frontier separating the SAF order with the paramagnetic disorder as shown in Figure 2. For completeness,  Figure 2  also includes the critical frontier obtained by MC simulatons for the particular case $\lambda_{1}=1$, $\lambda_{2}=0$ (square lattice), which lacks of first-order criticality \cite{viana2009}. In what effective-field calculations concerns, we obtain the critical frontier, which consists of  the transition temperature as a function of the external magnetic field, separating the SAF and P phases, by simultaneously solving the three transcendental equations Eqs. (\ref{15}), (\ref{16}) and (\ref{17}). The range of ratio $h=H/J_{x}$ determines the limits of second-order $(0<h<h_{t})$ and first-order $(h_{t}<h\leqslant2)$ phase-transition frontiers, where $(h_{t},k_{B}/J_{x}T_{t})$ is the tricritical point (TCP). From Landau theory, the TCP is calculated by the condition given by  $A_{1}(m,T_{t},H_{t})=1$, $A_{3}(m,T_{t},H_{t})=0$ and $A_{5}(m,T_{t},H_{t})>0$, obtaining in EFT-4 the following values:  $h_{t}=1.72$ and $k_{B}T_{t}/J_{x}=3.42$. The temperature variation of $m_{s}(T,H)$ for fixed reduced field $H/J_{x}$ present two type of behavior. The first for low-field the order parameter decrease to zero continuously as the reduced temperature approaches to the  critical point, the temperature ($H$ fixed) where $m_{s}=0$ is the second-order phase transition temperatute, $T_{c}(H)$. On the other hand, when we are at high-fied the temperature at which the order parameter make a jump discontinuity is the first-order phase transitions going to zero discontinuously at the point $T_{c}^{∗}(H)$. For h value greater than the upper limit of these field (i. e., $h>h_{c}=2.0$), the system exhibits no phase transition (the order parameter is null for all finite temperature, $T>0$)\\\\

In  order to obtain relevant critical points to get the critical frontier in MC simulations, we  estimate the  critical temperature for a given value of field by a finite-size scaling analysis. For instance, for $h=H/J_{x}=1.1$, Figure 3 shows how the critical temperature was located around the crossing point of $U_{4}-k_{B}T/J_{x}$ curves for sizes $L=16,32,64$, resulting in $k_{B}T_{c}/J_{x} \sim 4.14$. Note that  critical temperatures for lower fields  are  superestimated by the EFT-4 technique (see Figure 2).  The limit of the second-order frontier  was found by obtaining several $U_{4}-k_{B}T/J_{x}$  curves for different field values for a given lattice size. Consequently, we found the beginning of the first-order criticality after detecting a change in the behavior of the Binder Cumulant curve above a certain value of field. This is well exemplified in Figure 4, for two different field values $h=1.84$, $1.88$, for $L=30$. Figure 4a shows that the Binder Cumulant curve  already presents a clear first-order transition for $h = 1.88$, because of the sharply negative fluctuated peaks around the pseudocritical temperature. Moreover, Figure 4b also suggests this first-order criticality by the aparent Delta-Dirac form of the susceptivility, for  $h = 1.88$.  In order to confirm what is suggested in short lattice sizes, we draw upon the fact that around the critical temperature of a phase transition the susceptibility peaks behave as $\chi_{max} \sim L^{\theta}$, where $\theta=d$ ($d$ is the lattice dimension), for a first-order criticality, whereas  $\theta=\gamma/\nu$ for a second-order one. Consequently, Figure 5 shows a finite-size scaling of the susceptibility peak versus $L$, for sizes $L=15$, $20$, $30$ and $40$. 
For $h=1.84$, we estimated $\theta = 1.98(3)$, which is (within the error bar) an universal value \cite{gupta,flandau} for the second-order criticality of the three-dimensional Ising model. For $h = 1.88$, $\theta \sim 2.8$, which is close to $d=3$, which suggests a first-order transition. Therefore, the tricritical point must be in the interval $1.84 < h < 1.88$. On the other hand, the first-order curve must end at $h_{c}=2$.\\\\
We could estimate some points of the first-order frontier by a finite-size scaling method, however, in order to avoid finite-size problems to obtain well behavied curves for the specific heat or the susceptibility around  critical points belonging to the first-order frontier, we decided to estimate the corresponding critical temperatures for given fields by computing the free energy versus $\beta = 1/k_{B}T$, from Monte Carlo data of energy versus $\beta$, for a given size. For instance, in figure 6a it is shown the hysteresis effect   from energy data around the critical temperature obtained when cooling or heating the system too fast. This is because of huge intrinsic autocorrelation times in finite-size systems. We can improve the location of the critical temperature  by obtaining the associated free energies of the low- and high-temperature branches\cite{computersim}. Consequently, the crossing point of the  free energy branches  gives a good estimation of the critical temperature, inasmuch the stable phase has the lower free energy, as shown in figure 4b. \\\\
 The  first-order frontier in MC simulations does not present a reentrant behavior as EFT-4 frontier does. So, this reentrance seems to be an artifact of the effective-field approach.  

\section{Conclusions}

In summary, we investigated the anisotropic three-dimensional nearest-neighbor Ising model with competitive interations in an uniform longitudinal field. We obtained the phase diagram in the plane $h-k_{B}T/J_{x}$, where the critical frontier separates the SAF order with the paramagnetic disorder. For lower fields, the EFT-4 approach superestimates the critical temperatures obtained by MC simulations. A tricritical point was found for $1.84 < h < 1.88$, by a finite-size analysis. At low temperatures EFT-4 calculations show a reentrant first-order frontier, which does not appear by MC simulations. It suggests that improvements in treating correlations, or by increasing the cluster size in the Effective-Field approach, could correct this reentrant curve. At zero temperature, the critical field is exactly obtained, so  $h_{c}=2.0$. Our quantitative estimation for the tricritical point could be bettered by using larger lattice sizes with better Metropolis Monte Carlo techniques  like  Parallel Tempering. 

\textbf{ACKNOWLEDGEMENT}

We thank Professor David P. Landau for fruitful suggestions in the VI BMSP (Brazilian Meeting on Simulational Physics 2011).
This work was partially supported by CNPq (Edital Universal) and FAPEAM (Programa Primeiros Projetos - PPP) (Brazilian Research Agencies).

\vspace{0.4cm}
\begin{figure}[htbp]
\centering
\includegraphics[width=9.0cm,height=9.0cm]{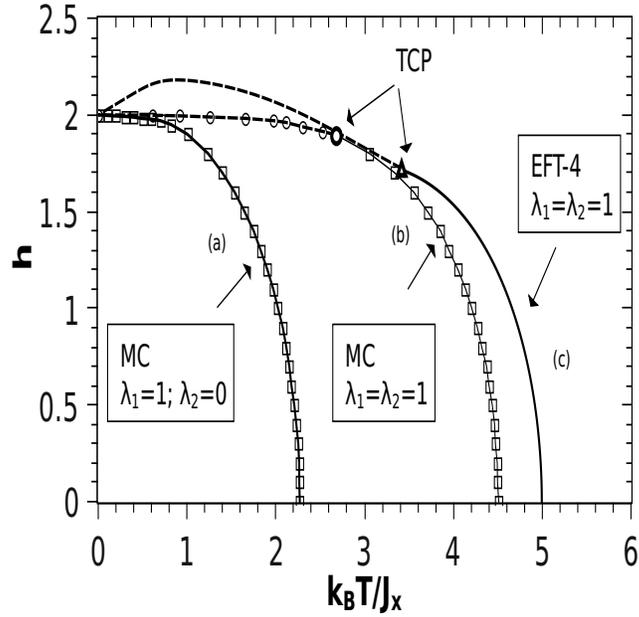}
\caption{Critical frontiers in the plane  $h-k_{B}T/J_{x}$ ($h=H/J_{x}$). These frontiers separate the colinear order (SAF) and the paramagnetic phase (P). The 
curve (a) was obtianed by  MC simulations  \cite{viana2009} for the present model implemented in  square lattices. The solid line corresponds to  second-order 
transitions. The curves (b) and (c) are our results obtainded by  MC and EFT-4 methods, respectively. The dashed lines corresponds to  first-order transitions appeared  for the present model implemented in  cubic lattices.} 
\label{saf}
\end{figure}

\vspace{0.4cm}
\begin{figure}[htbp]
\centering
\includegraphics[width=9.0cm,height=9.0cm]{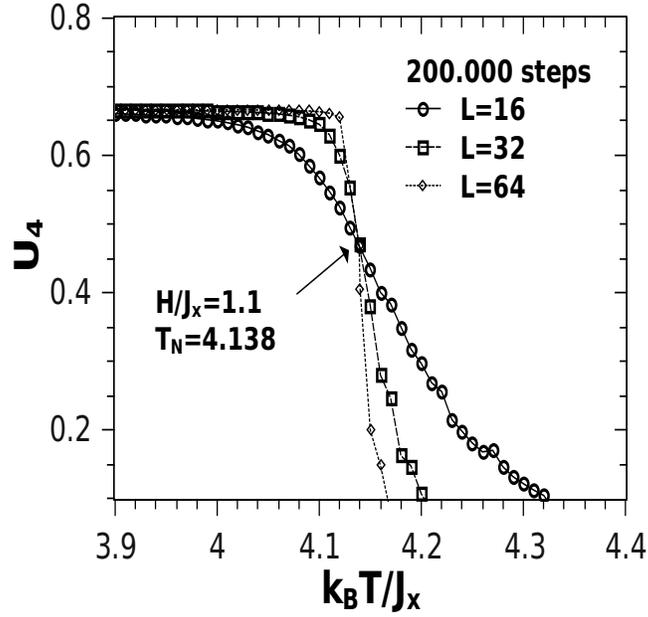}
\caption{The inset shows the fourth-order cumulant $U_{4}(L)$ for a particular field $h=H/J_{x}=1.1$ and system sizes $L=16,32$ and $64$.} 
\label{saf2}
\end{figure}

\begin{figure}[htbp]
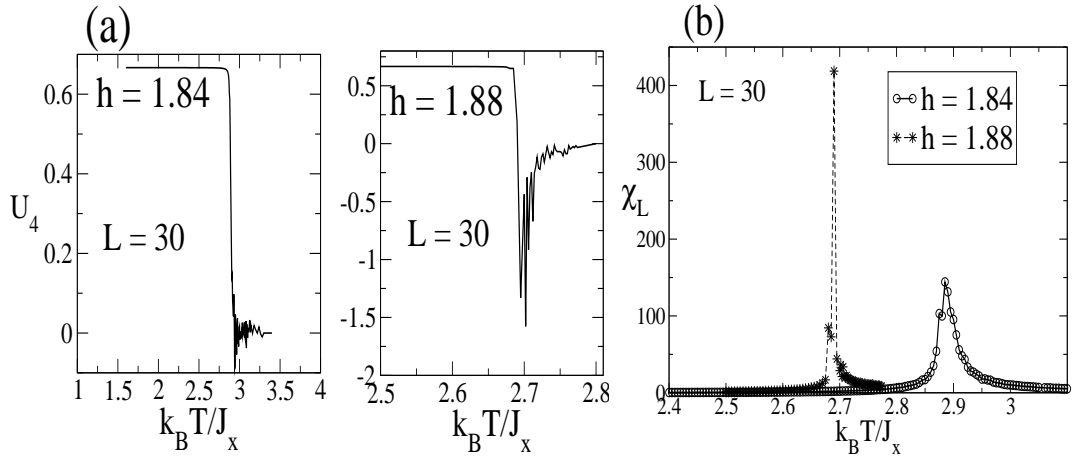

\centering
\includegraphics[width=8.0cm,height=6.0cm]{U4_ab.eps}
\includegraphics[width=6.0cm,height=6.0cm]{X_T.eps}
\caption{The two figures show, two different  phase transition orders at different values of the external  field, through  the Binder Cumulant (a), 
and through the susceptibility of the relevant order parameter (b), for the present model implemented in the sc for $L=30$.} 
\label{s1}
\end{figure}

\begin{figure}[htbp]
\centering
\includegraphics[width=9.0cm,height=9.0cm]{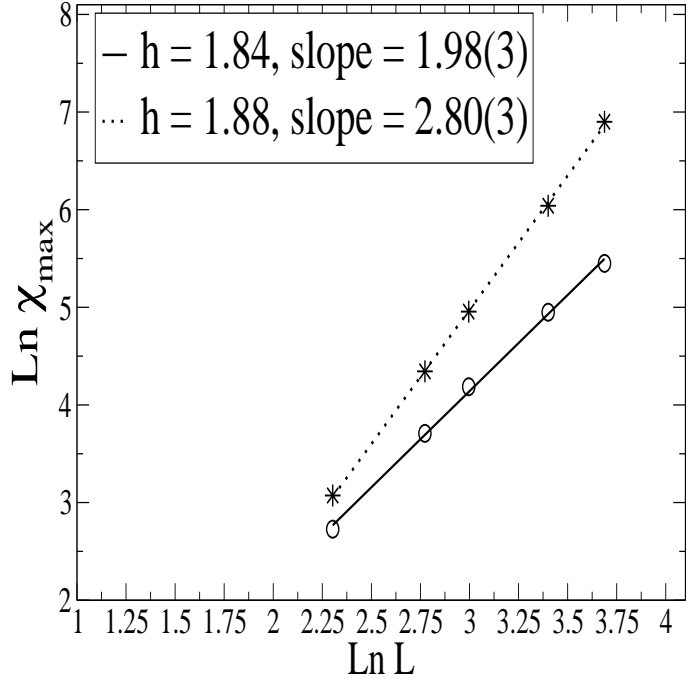}
\caption{Finite-size scaling of the susceptibility related to the relevant order parameter for two values of field, corresponding to  the present model 
implemented in the cubic lattice. The different estimated slopes suggest a tricritical point for  $1.84<h<1.88$.} 
\label{s2}
\end{figure}

\begin{figure}[htbp]
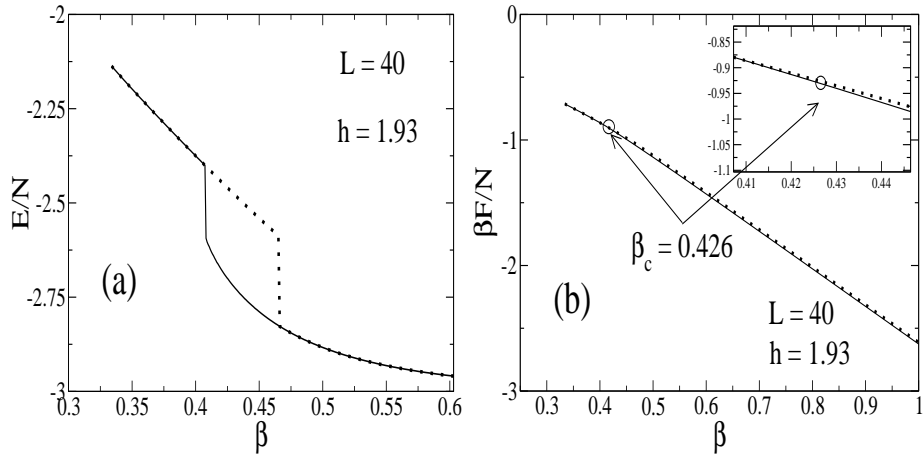

\centering
\includegraphics[width=6.0cm,height=6.0cm]{E_bL40H1y93.eps}
\includegraphics[width=6.0cm,height=6.0cm]{bF_bL40H1y93.eps}
\caption{(a) Heating and cooling Monte Carlo runs in $\beta=1/k_{B}T$, for the energy. (b) The associated free energy obtianed from the Monte Carlo data in (a), by thermodynamic integration. } 
\label{s3}
\end{figure}


\begin{thebibliography}{31}

\bibitem{dejongh} L. J. de Jongh et al., \textit{Physica} \textbf{58}, 277(1972) .

\bibitem{dejongh2} L. J. de Jongh et al., \textit{J. Appl. Phys.} \textbf{40}, 1363(1969).

\bibitem{bloembergen} P. Bloembergen et al., Proc. Int. Conf. Magn., Grenoble 1970 
\textit{J. Phys.}, suppl. no. 2-3 Tome \textbf{32},  879(1971).

\bibitem{dejongh3} L. J. de Jongh et al., Proc. Int. Conf. Magn., Grenoble 1970 
\textit{J. Phys.}, suppl. no. 2-3 Tome \textbf{32}, 880(1971).

\bibitem{miedema} A. R. Miedema, Proc. Int. Conf. Magn., Grenoble 1970 
\textit{J. Phys.}, suppl. no. 2-3 Tome \textbf{32},  305(1971).

\bibitem{mulder} C. A. M. Mulder et al., \textit{Physica B+C} \textbf{113},  380(1982).

\bibitem{heger} G. Heger et al., \textit{Solid State Commun.} \textbf{12}, 1157(1973).

\bibitem{domb} See, e.g., C. Domb, in Phase Transitions and Critical Phenomena, edited by C. Domb 
and M. S. Green (Academic, New York, 1974), Vol. 3.

\bibitem{wolf} W. P. Wolf, \textit{J. Phys.} (Paris) Suppl. \textbf{32}, C1(1971).

\bibitem{graim} T. Graim and D. P. Landau, \textit{Phys. Rev. B} \textbf{24},  5156(1981).  
\textit{Phys. Rev. Lett.} \textbf{68}, 9(1992).

\bibitem{stout} W. Stout and R. C. Chisholm, \textit{J. Chem. Phys.} \textbf{3C},  979(1962).  

\bibitem{hone} D. Hone, P. A. Montano and T. T. Tonegawa \textit{Phys. Rev. B} \textbf{12},  5141(1975).

\bibitem{sato} H. Sato, \textit{J. Phys. Chem. Solids} \textbf{19},  74(1961).

\bibitem{garrett} C. G. B. Garrett, \textit{J. Chem. Phys.} \textbf{19},  1154(1951).

\bibitem{ziman} J. M. Ziman, \textit{Proc. Phys. Soc.}, London, Sect. A \textbf{64},  1108(1951).

\bibitem{zukovic} M. Zukovic, A. Bobak and T. Idogaki \textit{J. Magn. Magn. Mater.} \textbf{192},  363(1999).

\bibitem{slotte} P. A. Slotte, \textit{J. Phys. C} \textbf{16},  2935(1983).

\bibitem{neto2004} Minos A. Neto and J. Ricardo de Sousa, \textit{Phys. Rev. B} \textbf{70},  224436(2004).

\bibitem{landau} D. P. Landau, \textit{Phys. Rev. B} \textbf{16},  4164(1977).

\bibitem{landau1976} D. P. Landau, \textit{Phys. Rev. B} \textbf{14},  255(1976). 

\bibitem{ferrenberg} For simple cubic lattice, see A. M. Ferrenberg and D. P. Landau, \textit{Phys. Rev. B} \textbf{44},  
5081(1991). 

\bibitem{bienenstock} A. Bienenstock and J. Lewis, \textit{Phys. Rev.} \textbf{160},  393(1967). 

\bibitem{viana2009} J. Roberto Viana, Minos A. Neto and J. Ricardo de Sousa \textit{Phys. Latt. A} 
\textbf{373},  2413(2009).

\bibitem{metropolis} M. Metropolis et al., \textit{J. Chem. Phys.} \textbf{21},  1087(1953); R. J. Glauber, 
J. Math. Phys. \textbf{4},  294(1963).

\bibitem{nemoto} K. Hukushima, and K. Nemoto \textit{J. Phys. Jpn.} \textbf{65},  1604(1996).

\bibitem{wolff} U. Wolff, \textit{Phys. Rev. Lett.} \textbf{62},  361(1989); R. H. Swendsen, and J. S. Wang, 
\textit{Phys. Rev. Lett.} \textbf{58},  86(1987).

\bibitem{berg} B. A. Berg, and T. Neuhaus, \textit{Phys. Lett. B} \textbf{267},  249(1991); 
\textit{Phys. Rev. Lett.} \textbf{68},  9(1992).

\bibitem{wanglandau} F. Wang, and D. P. Landau, \textit{Phys. Rev. Lett.} \textbf{86},  2050(2001).

\bibitem{onsager} L. Onsager, \textit{Phys. Rev.} \textbf{65},  117(1944).

\bibitem{fisher} M. E. Fisher, in \textit{Proceedings of the International Summer School Enrico Fermi, Course 51, 
Varenna 1970}, edited by M. S. Green (Academic, New York, 1971).

\bibitem{binder} K. Binder, \textit{Z. Phys. B}: Condens. Matter \textbf{43},  119(1981).

\bibitem{denise} Denise A. do Nascimento \textit{et al.}, \emph{J. Magn. Magn. Mater.}  \textbf{324},  2429(2012).

\bibitem{honmura} R. Honmura and T. Kaneyoshi, \emph{J. Phys. C} \textbf{12}, 3979(1979).

\bibitem{gupta} R. Gupta and P. Tamayo, \emph{IJMPC} \textbf{7} , 305(1996).

\bibitem{flandau} Alan M. Ferrenberg and D. P. Landau, \emph{Phys. Rev. B} \textbf{44}, 5081(1991). 

\bibitem{computersim} see pages 111-116 of W. Janke, Computer Simulations of Surfaces and Interfaces, edited by B. Dunweg, D. P. Landau, A. I. Milchev (NATO Science Series 2002). 


\vspace{10.0cm}
\end{thebibliography}
\end{document}